\documentclass[12pt]{article}

\usepackage{amssymb}
\usepackage{color}
\usepackage{epsfig,amssymb,amsfonts,amsmath,graphicx,dsfont,cite,xfrac}
\usepackage{authblk}
\usepackage{subfigure}
\definecolor{mygray}{gray}{0.5}

\usepackage{cite}
\usepackage[colorlinks=true,linkcolor=blue,citecolor=red]{hyperref}

\newcommand{\be}{\begin{equation}}
\newcommand{\ee}{\end{equation}}
\newcommand{\bea}{\begin{eqnarray}}
\newcommand{\eea}{\end{eqnarray}}

\title{Superpositions of bright and dark solitons supporting the creation of balanced gain-and-loss optical potentials 
}

\author[1]{Oscar Rosas-Ortiz\thanks{Corresponding author. E-mail: orosas@fis.cinvestav.mx (O. Rosas-Ortiz).}}
\author[2]{Sara Cruz~y~Cruz}

\affil[1]{\footnotesize Physics Department, Cinvestav, AP 14-740, 07000
M\'exico City, Mexico}

\affil[2]{\footnotesize Instituto Polit\'ecnico Nacional, UPIITA, Av. I.P.N. 2580, Col. La Laguna Ticom\'an, 07360 M\'exico City, Mexico}

\date{}
\begin{document}

\maketitle

\begin{abstract}
Bright and dark solitons of the cubic nonlinear Schr\"odinger equation are used to construct complex-valued potentials with all-real spectrum. The real part of these potentials is equal to the intensity of a bright soliton while their imaginary part is defined by the product of such soliton with its concomitant, a dark soliton. Considering light propagation in Kerr media, the real part of the potential refers to the self-focusing of the signal and the imaginary one provides the system with balanced gain-and-loss rates.
\end{abstract}


\section{Introduction}

Optical solitons are localized pulses that do not change shape as they propagate in nonlinear media \cite{Agr92,Tay92,Kiv03}. Dispersion and nonlinearity conspire to cancel the spatial dependence in the dynamics, which is usually described by nonlinear differential equations \cite{Lam80,Nov84}. In particular, the so-called bright solitons have a very practical presence in optics \cite{Tay92}. They are bound states of the cubic nonlinear Schr\"odinger equation, and exist because attractive nonlinearities are originated in the media by the Kerr effect \cite{Kiv98,Sul99}. Solutions for repulsive nonlinearities, known as dark optical solitons, are also available and useful \cite{Kiv98,Sul99}. Some recent applications of bright and dark solitons in optics and dispersive media can be found in, e.g. \cite{Kom15a,Kom15b,Bis17a,Bis17b,Yan18,Li18,Liu18,Yu18,Guo18,Zha18}. Of particular interest, the properties of optical solitons have been recently used to validate the parity-time symmetry in optics \cite{Rut10}, which means invariance under parity and time-reversal transformations in quantum mechanics \cite{Ben05}. 

The experimental proof proportioned in \cite{Rut10} is based on the formal equivalence between some dynamical equations in optics and the Schr\"odinger equation in quantum mechanics. Within such equivalence, a complex-valued quantum mechanical potential may be modeled through a complex-valued refractive-index $n(x) = n_R(x) + i n_I(x)$, which is actually realized in the laboratory. The gain and loss regions of the related optical medium are associated with the imaginary distribution $n_I(x)$, which may be chosen odd $n_I(-x)= - n_I(x)$ to balance the gain-and-loss rates. An even distribution $n_R(-x) = n_R(x)$ would guide the signal along the propagation direction that is transversal to $x$.

On the other hand, it has been shown that exactly solvable complex-valued potentials with all-real spectra are available in quantum mechanics \cite{Ros15,Zel16,Jai17,Ros18,Bla19,Zel01}. The approach used to construct such potentials is based on the properties of the Riccati equation in the complex domain \cite{Hil97,Sch18}, the Ermakov equation \cite{Erm80}, and the Darboux method \cite{Dar82}. The Riccati and Ermakov equations are nonlinear, of first and second order respectively, with well known methods of solution \cite{Hil97,Erm80}. In turn, the Darboux method is very useful to construct new exactly solvable potentials by pairing their spectra with the energies of a given initial potential, the spectral problem of which is already solved \cite{Mie04}. This method finds immediate applications in research topics like soliton theory \cite{Rog02} and supersymmetric quantum mechanics \cite{Mie04,Coo01}. 

The present work is addressed to construct complex-valued potentials that can be modeled in optics by Kerr media with balanced gain-and-loss rates. Our principal interest is to provide such potentials with the appropriate nonlinearities to guide self-focusing waves. The subject is to find a Darboux-deformed potential that can be expressed as the complex-valued function $V= -(\vartheta^2 + i \vartheta_x)$, with $\vartheta$ a soliton-like solution of the appropriate nonlinear differential equation (to be determined), and $\vartheta_x$ the corresponding $x$-derivative. The proposal is motivated by the fact that the combination $\vartheta^2 + i \vartheta_x$ has been successfully used as seed of solutions for nonlinear differential equations \cite{Lam80} and would provide, in our case, the nonlinear potential we look for.

Given two potentials, $V_0$ and $V$, paired in their energy spectrum by a Darboux transformation, in this work we show that $V$ can be written in the form $V= -(\vartheta^2 + i \vartheta_x)$ whenever $V_0$ is the free-particle potential and $\vartheta$ is a solution of the time-independent Gross--Pitaevskii nonlinear equation \cite{Gro61,Pit61}. Extending the above results to $z$-parameterized wave amplitudes $\Theta$, with $z$ the distance along the propagation direction (which is transversal to the position variable $x$), we show that $\Theta$ is a bright soliton-like solution of the cubic Schr\"odinger equation \cite{Kiv98,Sul99} that can be factorized as the product of $\vartheta$ with a phase that depends on $z$. In this form, the real part of $V$ is equal to the intensity of the bright soliton while the imaginary part results from the product of such soliton with its concomitant, a dark soliton.

The rest of the paper is organized as follows. In Section~\ref{model} we formulate the problem to solve by revisiting first the Darboux construction of complex-valued potentials with all-real spectra. The conditions that must be satisfied by the initial potential $V_0$ and the functions $\vartheta$ used to get $V =-(\vartheta^2 + i \vartheta_x)$ are investigated in Section~\ref{GPE}. The extension to time-dependent soliton-like solutions $\Theta$ is developed in Section~\ref{time}. Concrete expressions for $V$ are provided in Section~\ref{optical}. Conclusions are given in Section~\ref{concluye}. Appendix~\ref{ApA} includes detailed calculations of some concrete expressions appearing in the paper.

\section{Problem formulation}
\label{model}

Using the Darboux approach \cite{Dar82}, the stationary one-dimensional Schr\"odinger equations,
\be
-\psi_{xx} +V \psi =  k^2 \psi
\label{nl1}
\ee
and
\be
-\varphi_{xx} +V_0 \varphi =  k^2 \varphi,
\label{nl2}
\ee
can be intertwined through the relationship 
\be
V = V_0 + 2 \beta_x,  \quad \psi= \varphi_x + \beta \varphi,
\label{darboux}
\ee
with $\beta$ a solution of the nonlinear Riccati equation 
\be
-\beta_x +\beta^2 =V_0 -\epsilon.
\label{ricatti}
\ee
Assuming that $V_0$ is a real-valued function such that Eq.~(\ref{nl2}) is integrable in $\operatorname{Dom} V_0 = (a_1,a_2) \subseteq \mathbb R$, with the real eigenvalues $E=k^2$, one can construct a complex-valued function $V$ such that Eq.~(\ref{nl1}) is integrable in $\operatorname{Dom} V = \operatorname{Dom} V_0$, with the same energies as $V_0$, plus an additional real eigenvalue $\epsilon$ \cite{Ros15}. Indeed, given $\epsilon \in \mathbb R$, a complex-valued solution $\beta = \beta_R + i \beta_I$ of (\ref{ricatti}) must satisfy the coupled system 
\be
-\beta_{R x} + \beta_R^2 - \beta_I^2 + \epsilon -V_0 =0,
\label{rica1}
\ee
 \be
 -\beta_{I x} + 2 \beta_I \beta_R =0.
 \label{rica2}
 \ee
 The straightforward calculation shows that $\beta$ is parameterized by a real number $\lambda$ \cite{Ros15}:
\be
\beta = -\frac{\alpha_x}{\alpha} + i\frac{\lambda}{\alpha^2}, \quad \lambda \in \mathbb R,
\label{beta}
\ee
where $\alpha$ satisfies the Ermakov equation
\be
\alpha_{xx} - (V_0 - \epsilon) \alpha = \frac{\lambda^2}{\alpha^3},
\label{Ermakov}
\ee
and is given by 
\be
\alpha(x) =  \left[ av^2(x) + b v(x) u(x) + c u^2(x) \right]^{1/2}.
\label{alpha}
\ee
The function $\alpha$ is positive semidefinite in $\mbox{Dom}V_0$ if the parameters $\{a,b,c\}$ are all real and such that $4ac- b^2 = 4 (\lambda/w_0)^2$ \cite{Ros15,Bla19}. The functions $u$ and $v$ are two linearly independent solutions of Eq.~(\ref{nl2}) for $k^2=\epsilon$, with $w_0 = W(u,v)$ the corresponding Wronskian.

The potential defined by the Darboux transformation (\ref{darboux}) is therefore parameterized by the real number $\lambda$ as follows
\be
V_{\lambda} = V_0 + 2 \left(\beta_{Rx} + i \beta_{I x} \right) = V_0 -2 \left( \frac{\alpha_x}{\alpha} \right)_x -i 4 \lambda \frac{\alpha_x}{\alpha^3}, \quad \lambda \in \mathbb R.
\label{potcomp}
\ee 
Additionally, it may be shown \cite{Jai17} that the appropriate  $\alpha$-functions provide potentials $V_{\lambda}$ satisfying the {\em condition of zero total area},
\be
\int_{\operatorname{Dom} V_0} \mbox{Im} V_{\lambda} (x) dx =0,
\label{zero}
\ee
so that the total probability is conserved. The latter means that potentials (\ref{potcomp}) can be addressed to represent quantum systems with balanced gain (acceptor) and loss (donor) profile \cite{Ele17}.

We look for complex-valued potentials (\ref{potcomp}) that can be written in the form
\be
V_{\lambda}= - (\vartheta^2 + i \vartheta_x),
\label{pot1}
\ee
where $\vartheta$ is a soliton-like solution of a nonlinear differential equation to be determined.

The usefulness of the approach we are going to develop is twofold. On the one hand, the relationship between (\ref{potcomp}) and (\ref{pot1}) permits to connect the balanced gain-and-loss profile of $V_{\lambda}$ with the soliton-like behavior of waves propagating in dispersive media. On the other hand, such relationship supplies a meaning for the real and imaginary parts of the $\beta$-function, a subject which is rarely discussed in the literature of the Darboux transformation.

\subsection{The time-independent Gross--Pitaevskii equation}
\label{GPE}

In this section we identify the nonlinear differential equation obeyed by the soliton-like function $\vartheta$ in the definition of $V_{\lambda}$ supplied by Eq.~(\ref{pot1}). The latter implies a concrete form of the initial potentials $V_0$, as we are going to see.

First, comparing (\ref{potcomp}) with (\ref{pot1}) we arrive at the system
\be
V_0 + 2 \beta_{R x} =- \vartheta^2, \qquad 2\beta_{I x} =- \vartheta_x.
\label{system}
\ee
Integrating the last of the above equations gives
\be
\vartheta = -2 \beta_I + \vartheta_0,
\label{teta1}
\ee
with $\vartheta_0$ an integration constant. The combination of (\ref{teta1}) with (\ref{rica2}) produces 
\be
\vartheta_x = 2 (\vartheta -\vartheta_0) \beta_R.
\label{teta2}
\ee
Then, the real and imaginary parts of $\beta$ lead to the following system
\be
\beta_R = \frac{\vartheta_x}{ 2 (\vartheta - \vartheta_0) } = -\frac{\alpha_x}{\alpha},  \qquad \beta_I = - \left( \frac{\vartheta - \vartheta_0}{2} \right) = \frac{\lambda}{\alpha^2}.
\label{betas}
\ee
Solving (\ref{betas}) we obtain\footnote{The constant arising from the integration of the equation associated with $\beta_R$ has been fixed as $-2\lambda$ for consistency.} a relationship  between $\vartheta$ and $\alpha$:
\be
\vartheta = -\frac{2 \lambda}{\alpha^2} + \vartheta_0.
\label{sol}
\ee
The above result allows to rewrite (\ref{potcomp}) in the form (\ref{pot1}). However, it is still open the determination of the nonlinear equation that must be satisfied by the $\vartheta$ function (\ref{sol}). To fill such gap let us introduce (\ref{teta1}) into the equation for $\beta_R$ in (\ref{system}). After using Eq.~(\ref{rica1}) we obtain
\be
2 (\beta^2_R + \beta^2_I + \epsilon) -V_0 = \vartheta_0 (4 \beta_I -\vartheta_0).
\ee
Without loss of generality we make $\vartheta_0 =0$. Then, the above equation is reduced to the constraint
\be
\vert \beta \vert^2 = \tfrac12 V_0 -\epsilon.
\label{modbeta}
\ee
As $\vert \beta \vert \geq 0$ we immediately have $V_0 \geq 2 \epsilon$. Besides, from (\ref{betas}) we realize that (\ref{modbeta}) produces the nonlinear differential equation 
\be
\vartheta_x^2 + (4\epsilon - 2 V_0) \vartheta^2 + \vartheta^4 = 0,
\label{tetita}
\ee
which defines the analytic form of $\vartheta$. 

The next step is to determine whether or not the function $\vartheta$ features a soliton profile. With this aim notice that the derivative of (\ref{teta2}), after using (\ref{system}), (\ref{betas}) and (\ref{modbeta}), gives
\be
-\vartheta_{xx} + (V_0  - 2 \vartheta^2 ) \vartheta = 4  \epsilon \vartheta.
\label{teta4}
\ee
From (\ref{alpha}) and (\ref{sol}) we know that $\vartheta$ is a real-valued function, so we can rewrite (\ref{teta4}) as follows
\be
-\vartheta_{xx} + \left( V_0 - 2 \vert \vartheta \vert^2 \right) \vartheta = 4  \epsilon \vartheta.
\label{teta4b}
\ee
The spectral problem (\ref{teta4b}) is named after Gross \cite{Gro61} and Pitaevskii \cite{Pit61}, although it is currently known as the time-independent Gross--Pitaevskii equation. 

Clearly, the non-linear differential equations (\ref{tetita}) and (\ref{teta4})-(\ref{teta4b}) must be consistent. The derivative of Eq.~(\ref{tetita}), after assuming that (\ref{teta4}) is satisfied, leads to the condition $V_0 \vartheta= \operatorname{const}$ (see Appendix~\ref{ApA}). As $V_0$ is an input in our algorithm, for nontrivial functions $\vartheta$ we have to demand $V_0=0$. That is, only the free-particle potential may give rise to the complex-valued potentials $V_{\lambda}$ we are looking for. The same conclusion is obtained by using the relationship between $\beta$ and $\alpha$ through the Ermakov equation (\ref{Ermakov}). Indeed, it may be proven that the time-independent Gross--Pitaevskii equation (\ref{teta4b}) and the Ermakov one (\ref{Ermakov}) are interrelated by the transformation (\ref{sol}) whenever $V_0=0$ (see details in Appendix~\ref{ApA}).

\subsection{Extension to time-dependent soliton-like solutions}
\label{time}

We can associate the solutions of the time-independent Gross--Pitaevskii equation (\ref{teta4b}) with a $z$-parameterized function $\Theta(x;z)$, factorized in the form
\be
\Theta(x;z) = \vartheta(x) \exp ( - i 4 \epsilon z + \xi_0 ),
\label{time2}
\ee
that solves the linear differential equation 
\be
i \Theta_z = 4 \epsilon \Theta.
\label{time1}
\ee
In the above expressions $\xi_0$ is an integration constant and $z$ is a real parameter. Combining (\ref{teta4b}) and  (\ref{time1}) one obtains
\be
-\Theta_{xx} + \left( V_0  - 2 \vert \Theta \vert^2  \right) \Theta = i \Theta_z.
\label{teta5}
\ee
The latter is called time-dependent Gross--Pitaevskii equation (GPE for short), mainly when the propagation parameter $z$ is treated as the evolution variable. In analogy with the Schr\"odinger equation, $V_0$ is an external potential and the nonlinear term $-2\vert \Theta \vert^2$ represents an attractive interaction that is proportional to the local density $\vert \Theta \vert^2 = \vert \vartheta \vert^2$. The GPE is a powerful tool to study Bose-Einstein condensates (BEC) in the mean-field  approximation \cite{Rog13}, where the nonlinearity represents an effective potential to which each atom is subject because its interaction with all other particles, and $\vert \vartheta \vert^2$ stands for the atomic density. Within such approach the external potential $V_0$ produces the BEC confinement and may adopt different forms \cite{Kev08}. However, as a matter of fact, the GPE (\ref{teta5}) cannot be solved analytically for arbitrary $V_0$. Particular examples include periodic potentials $V_0(x+L)= V_0(x)$ with period $L$ for which the Bloch theory \cite{Koh59} gives rise to discrete solitons \cite{Tro01}. 

The simplest case in which Eq.~(\ref{teta5}) is exactly solvable corresponds to take $V_0$ as the free-particle potential, which leads to the cubic nonlinear Schr\"odinger equation (NLSE),
\be
- \Theta_{xx} - 2 \vert \Theta \vert^2 \Theta = i\Theta_z.
\label{nlse}
\ee
Eq.~(\ref{nlse}) is useful to describe the dynamics of complex field envelopes in nonlinear dispersive media \cite{Sul99}, as well as the paraxial approximation of light propagation in Kerr media \cite{Kiv03}. In the latter case, the propagation parameter $z$ refers to the distance along the beam and the variable $x$ stands for the direction transverse to the propagation. Therefore,  $\Theta$ is the normalized amplitude of the electric field envelope describing the pulse. The nonlinearity $-2\vert \Theta \vert^2$ is due to the Kerr effect and represents the refractive index, its effect on the light rays increases with the light intensity $\vert \Theta \vert^2$ and leads to the self-focussing of the beam \cite{Tay92}, Ch.1 (see also \cite{Kiv03} and \cite{Sul99}). In counterposition to the GPE (\ref{teta5}), the NLSE (\ref{nlse}) is exactly integrable in the inverse scattering approach \cite{Zak71}. It possesses localized solutions representing `bright' solitons while its counterpart, constructed with a repulsive nonlinearity $+2\vert \Theta \vert^2$, includes localized `dark' pulses \cite{Kiv98}.
 
In the previous section we have proven that only the free-particle potential produces complex-valued potentials $V_{\lambda}$ with the profile (\ref{pot1}). The functions $\vartheta$ defining $V_{\lambda}$ are therefore solutions of the time-independent GP equation (\ref{teta4})-(\ref{teta4b}) for $V_0=0$. At the same time $\vartheta$ defines solutions to the cubic nonlinear Schr\"odinger equation (\ref{nlse}) in the factorized form (\ref{time2}), so it can be considered a stationary solution of the problem.

We now face the solving of Eq.~(\ref{teta4}) for $V_0=0$, which is easily achieved by using the Darboux formalism of Section~\ref{model}, as we are going to see.

\section{Balanced gain-and-loss optical potentials}
\label{optical}

\begin{figure}[h!]
\centering
\subfigure[$\operatorname{Re} \beta (x)$]{\includegraphics[width=0.3\textwidth]{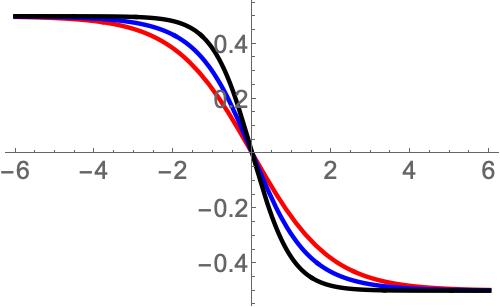}} 
\hspace{2ex}
\subfigure[$\operatorname{Im} \beta(x)$]{\includegraphics[width=0.3\textwidth]{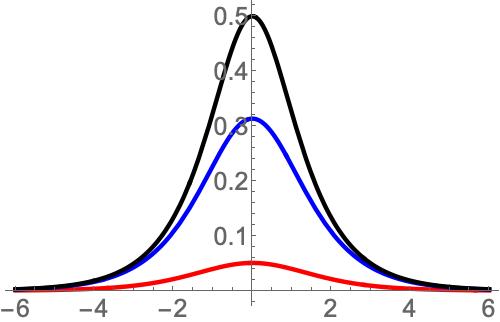}}

\caption{\footnotesize 
(Color online) The real and imaginary parts of the $\beta$-function defined in Eq.~(\ref{clave}) for $\kappa=1$, with $\lambda =0.1$ (red), $\lambda=0.45$ (blue), and $\lambda =0.5$ (black). The functions $\operatorname{Re} \beta$ and $\operatorname{Im} \beta$ are respectively odd and even with respect to the position variable $x$ (horizontal axis). However, they are even and odd under the transformation $\lambda \rightarrow -\lambda$.
}
\label{Figbeta}
\end{figure}

Considering the free-particle potential $V_0=0$, let us take $v=e^{-ikx}$ and $u= e^{ikx}$ as the fundamental solutions of Eq.~(\ref{nl2}), with $k^2 = \epsilon$ and $w_0 = -2ik$. To get regular potentials $V_{\lambda}$ we make $k=i \frac{\kappa}{2}$, with $\kappa >0$. Then, we arrive at the hyperbolic expressions \cite{Ros15}:
\be
\alpha(x) = \sqrt{ \cosh (\kappa x) + \sigma}, \quad \beta(x) = \frac{ -\tfrac{\kappa}{2} \sinh (\kappa x) + i \lambda}{\cosh(\kappa x) + \sigma}, \quad \sigma = \sqrt{1 - 4 \left( \tfrac{\lambda}{\kappa} \right)^2},
\label{clave}
\ee
where we have made $a=c=1/2$ for simplicity. Remark that the $\alpha$-function is positive semidefinite for $-\tfrac{\kappa}{2} \leq \lambda \leq \tfrac{\kappa}{2}$. Figure~\ref{Figbeta} shows the behavior of the real and imaginary parts of the $\beta$-function for $\kappa=1$ and three different values of $\lambda$ in the interval $[-1/2,1/2]$. Notice that $\operatorname{Re} \beta= \beta_R$ and $\operatorname{Im} \beta= \beta_I$ are respectively odd and even functions with respect to the position variable $x$. Quite interestingly, they are respectively even and odd under the change $\lambda \rightarrow -\lambda$.

Using (\ref{clave}) and (\ref{potcomp}) one arrives at the complex-valued potentials \cite{Ros15}:
\be
V_{\lambda}(x)= - \frac{\kappa^2 [1 + \sigma \cosh(\kappa x) ] + i 2 \lambda \kappa \sinh (\kappa x)}{(\cosh(\kappa x) +\sigma)^2}, \quad -\tfrac{\kappa}{2} \leq \lambda \leq \tfrac{\kappa}{2}.
\label{family1}
\ee
For $\lambda=0$ the potential (\ref{family1}) is real-valued and coincides with the conventional supersymmetric partner of the free-particle potential \cite{Coo01,Dia99,Mie00}. For $\lambda \neq 0$, it defines a family of non-Hermitian supersymmetric partners of $V_0=0$ \cite{Ros15}, and may be classified in the Scarf I-hyperbolic type \cite{Lev00}. 

\begin{figure}[h!]
\centering
\subfigure[$\operatorname{Re}V_{\lambda}(x)$]{\includegraphics[width=0.3\textwidth]{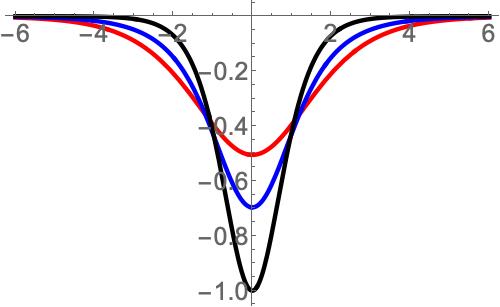}} 
\hspace{2ex}
\subfigure[$\operatorname{Im} V_{\lambda}(x)$]{\includegraphics[width=0.3\textwidth]{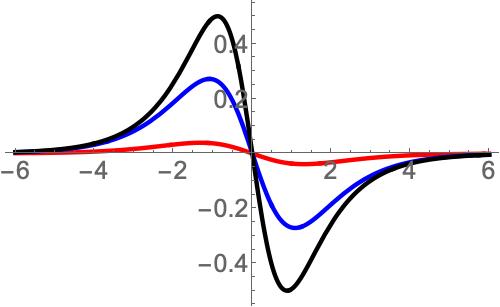}}

\caption{\footnotesize 
(Color online) The real and imaginary parts of the complex-valued potential $V_{\lambda}$ defined in Eq.~(\ref{family1}) for the parameters used in Figure~\ref{Figbeta}. The functions $\operatorname{Re} V_{\lambda}$ and $\operatorname{Im} V_{\lambda}$ are respectively even and odd with respect to the position variable $x$, which makes $V_{\lambda}$ invariant under the parity-time symmetry, $V_{\lambda}(x)=V_{\lambda}^*(-x)$. They are also even and odd under the transformation $\lambda \rightarrow -\lambda$, so that $V_{\lambda}(x) = V_{-\lambda}^*(x)$.
}
\label{FigPot}
\end{figure}

The symmetry properties of $\beta_R$ and $\beta_I$ provide the potentials (\ref{family1}) with the so-called parity-time symmetry \cite{Ben05}. That is, the potentials defined in (\ref{family1}) satisfy the condition $V_{\lambda}(x)=V_{\lambda}^*(-x)$, with $z^*$ the complex-conjugate of $z\in \mathbb C$, see Figure~\ref{FigPot}. In the present case, such a symmetry is a consequence of making $a=c$ in the construction of the $\alpha$-function, but it can be broken for the more general case $a\neq c$ (see the discussion on the matter given in \cite{Zel01}). 

An additional symmetry of the family of potentials (\ref{family1}) arises under the transformation $\lambda \rightarrow - \lambda$, for which $V_{\lambda}(x) = V_{-\lambda}^*(x)$, see Figure~\ref{FigPot}. The latter is a consequence of the fact that $\operatorname{Re} V_{\lambda}$ and $\operatorname{Im} V_{\lambda}$ are even and odd with respect to the parameter $\lambda$. Notice also that the function $\operatorname{Im} V_{\lambda}$ satisfies the condition of zero total area (\ref{zero}), which ensures the balanced gain-and-loss profile we are looking for.

\begin{figure}[h!]
\centering
\includegraphics[width=0.3\textwidth]{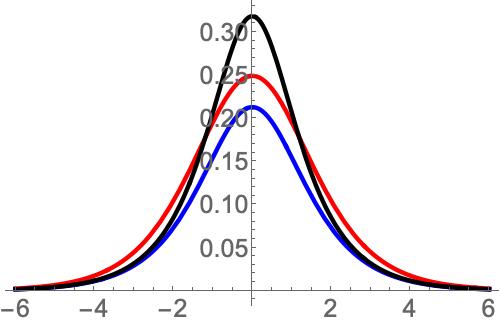}
\caption{\footnotesize 
(Color online) The probability density $\vert \psi_0(x;\lambda) \vert^2$ of the wave-function defined in Eq.~(\ref{groundstate}) for the parameters used in Figure~\ref{Figbeta}. In each case, the probability distribution corresponds to the single bound state allowed by the potentials depicted in Figure~\ref{FigPot}. In all cases the related energy is $E_0=-\frac14$.
}
\label{Figground}
\end{figure}

On the other hand, it may be shown that only the real eigenvalue $E=\epsilon =-\frac{\kappa^2}{4}$ permits a normalizable solution of the Schr\"odinger equation (\ref{nl1}), which is given by
\be
\psi_0(x; \lambda)= \frac{c_0}{\alpha(x)} \exp\left\{ i \operatorname{arctanh} \left[ \tfrac{\kappa(1-\sigma)}{2 \lambda} \tanh \left( \frac{\kappa x}{2} \right) \right] \right\},
\label{groundstate}
\ee
with $c_0$ the normalization constant. We have depicted the probability density $\vert \psi_0(x;\lambda) \vert^2$ in Figure~\ref{Figground} for the parameters used in Figs.~\ref{Figbeta} and \ref{FigPot}.

As we can see, all the complex-valued potentials belonging to the $\lambda$-parameterized family (\ref{family1}) share the same single-point discrete spectrum $\operatorname{Sp} ({H}_{\lambda })=\{E_0=-\frac{{\kappa }^{2}}{4}\}$ and have a balanced gain-and-loss profile. In the next section we show that, for the appropriate selection of the parameter $\lambda$, this family permits also the structure defined in Eq.~(\ref{pot1}).

\subsection{Optical soliton engineering}
\label{NLSE}

From Eq.~(\ref{sol}), using (\ref{clave}) we may write (it must be recalled that we have made $\vartheta_0=0$):
\be
\vartheta (x; \lambda) = - \frac{2\lambda}{\cosh(\kappa x) + \sigma}, \quad -\tfrac{\kappa}{2} \leq \lambda \leq \tfrac{\kappa}{2}.
\label{sol2}
\ee
Among the family of solutions (\ref{sol2}) we have to select those that also satisfy (\ref{tetita}) for the free-particle potential $V_0=0$. To find an expression for $\vartheta$ let us divide the nonlinear equation (\ref{tetita}) by $\vartheta^4$. After introducing $y =- \vartheta^{-1}$  we arrive at the nonlinear equation
\be
y_x^2 -  \kappa y^2 +1=0,
\label{tetita2}
\ee
where we have used $\epsilon = -\kappa^2/4$. The straightforward calculation shows that the functions $y = \pm \kappa^{-1} \cosh [ \kappa (x+x_0) ]$ solve Eq.~(\ref{tetita2}), with $x_0$ an integration constant. Comparing with (\ref{sol2}) we immediately recognize that the parameter values $\lambda = \pm \tfrac{\kappa}{2}$ provide the solutions of (\ref{tetita}) we are looking for. Hereafter we fix $\lambda=- \tfrac{\kappa}{2}$  and write $\vartheta(x, \lambda= -\frac{\kappa}{2}) \equiv \vartheta(x)$  for short. The transformation $\lambda \rightarrow - \lambda$ permits to recover the case $\lambda = \tfrac{\kappa}{2}$, with the symmetries discussed in Section~\ref{GPE}. Accordingly,  the $\beta$-function introduced in (\ref{clave}) acquires the form
\be
\beta(x) = -\frac{\kappa}{2} \tanh (\kappa x) - i \frac{\kappa}{2 \cosh(\kappa x)},
\label{betanew}
\ee
and the complex-valued potential (\ref{family1}) is reduced to the expression
\be
V(x) \equiv V_{\lambda = -\tfrac{\kappa}{2}}(x) = -\frac{\kappa^2}{\cosh^2 (\kappa x)} \left[ 1- i  \sinh(\kappa x) \right].
\label{mipot}
\ee
Interestingly, the potential (\ref{mipot}) has been implemented as the external field in the GPE \cite{Mus08}. Besides, the existence and stability of solitons in these potentials, with self-focusing and self-defocusing nonlinear cases, has been  recently investigated in e.g. \cite{Mid14,Tso14,Che14}. 

\begin{figure}[h!]

\centering
\includegraphics[width=0.3\textwidth]{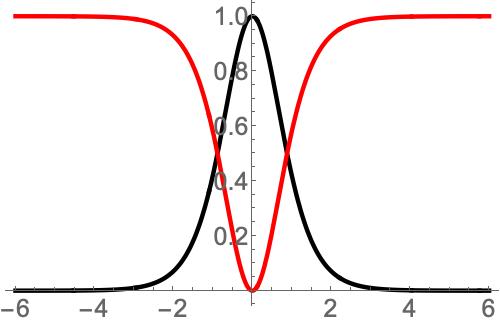}

\caption{\footnotesize 
(Color online) The local densities of the bright (black) and dark (red) solitons, defined respectively in Eqs.~(\ref{tetaz2}) and (\ref{dark}), are $z$-independent.}
\label{Figinten}
\end{figure}

On the other hand, the $z$-parameterized function (\ref{time2}) is in this case given by
\be
\Theta (x;z) =  \frac{\kappa e^{ ( i\kappa^2 z + \xi_0) } }{\cosh \left[ \kappa (x+x_0) \right]}.
\label{tetaz}
\ee
Without loss of generality we make $x_0 = \xi_0 =0$ to reduce (\ref{tetaz}) to the conventional form of the fundamental bright soliton 
\be
\Theta (x;z) =  \frac{\kappa e^{  i\kappa^2 z  } }{\cosh ( \kappa x) }.
\label{tetaz2}
\ee
The local density $\vert \Theta(x;z) \vert^2 = \vert \vartheta(x) \vert^2= \kappa^2/\cosh^2(\kappa x)$ does not change shape as the pulse propagates along the $z$-axis (see Figure~\ref{Figinten}), as it is expected from the balanced relationship between nonlinearity and dispersion in soliton profiles \cite{Lam80}. 

The imaginary part of the $\beta$-function in (\ref{betanew}) can be now expressed in terms of the bright soliton solution (\ref{tetaz2}). That is,
\be
\beta_I (x) = - \tfrac12 \left. \Theta (x;z) \right\vert_{z=0} =  -\frac12 \vartheta(x).
\label{bsoliton}
\ee
In turn, to connect the real part of $\beta$ with a soliton-like solution, let us introduce the $z$-parameterized function 
\be
\widetilde\Theta(x; z) = \widetilde\vartheta(x) \exp \left( i\kappa^2 z \right), \quad \widetilde\vartheta(x) = \kappa \tanh (\kappa x),
\label{dark}
\ee
which also satisfies Eq.~(\ref{time1}). We immediately recognize (\ref{dark}) as the fundamental dark soliton solution of Eq.~(\ref{nlse}), where the attractive nonlinearity $-2 \vert \Theta \vert^2$ is replaced by the repulsive one $+2 \vert \widetilde\Theta \vert^2$. As it is shown in Figure~\ref{Figinten}, the local density $\vert \widetilde\Theta(x;z) \vert^2 = \vert \widetilde \vartheta (x) \vert^2 = \kappa^2 \tanh^2(\kappa x)$ does not change shape along the $z$-axis. In this case, we may write
\be
\beta_R(x)= - \frac12 \left. \widetilde\Theta(x;z) \right\vert_{z=0} = - \frac12 \widetilde\vartheta(x).
\label{dsoliton2}
\ee
Therefore 
\be
\beta = - \tfrac12  \left( \widetilde\Theta + i \Theta \right)_{z=0} = -\tfrac12 ( \widetilde\vartheta + i \vartheta ).
\label{newbeta}
\ee 

To complete our program notice that, using $\vartheta_x = -\vartheta \widetilde\vartheta$, the potential (\ref{mipot})  acquires the profile (\ref{pot1}) in natural form. Taking full advantage of such a property let us rewrite (\ref{mipot}) as follows
\be
V (x)= -  \vartheta^2  + i \vartheta \widetilde\vartheta = -\vert \Theta (x;z) \vert^2 + i \Theta(x;z) \widetilde\Theta^*(x;z).
\label{pt1}
\ee
That is, the real part of the potential (\ref{mipot}) is defined by the bright soliton intensity (with no dependence on $z$), while the imaginary part results from the product of the bright and dark solitons, the latter is such that the $z$-dependence of the factors is cancelled. The nonlinearity $\vert \Theta (x;z) \vert^2$ provided by the real part of the potential increases with the intensity of the propagating wave and induces self-focusing. The effect of the imaginary part of the potential on the signal is also proportional to the intensity but it is modulated by the sign of the $\sinh(\kappa x)$ function. The latter produces a balanced gain-and-loss profile in the signal intensity overall the $x$-axis.

\begin{figure}[h!]

\centering
\includegraphics[width=0.3\textwidth]{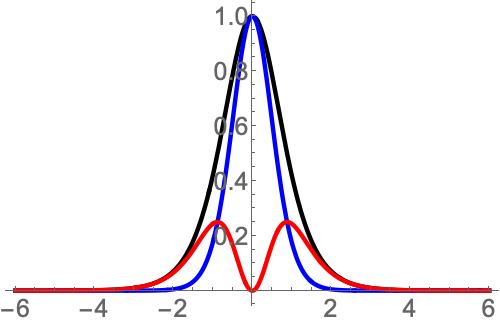}

\caption{\footnotesize 
(Color online) Total intensity (black) of potential (\ref{pt1}). The contribution of the real part (blue) is stronger than the imaginary one (red). }
\label{Fig4}
\end{figure}

At the present state one would wonder about the possibility of finding complex eigenvalues associated to potential (\ref{pt1}), equivalently (\ref{mipot}). In this respect we would like to emphasize that such potential is a family member of the parity-time-invariant potentials already studied in \cite{Ahm01} to show that non-Hermitian Hamiltonians have both real and complex discrete spectrum. The model discussed in \cite{Ahm01} includes different global factors for $V_R$ and $V_I$, and investigates whether the eigenvalues are real or complex in terms of such parameters. There, it is conjectured that ``when the real part of the parity-time-invariant potential is stronger than its imaginary part, the eigenspectrum will be real, and they will be mixed (real and complex) otherwise''. As our model considers the same global factor for $V_R$ and $V_I$, namely $\kappa^2$, the above conjecture is automatically verified (see Fig.~\ref{Fig4}), so that no complex eigenvalues are expected.

\subsection{Pulse propagation}

In this section we look for a way to provide the $\beta$-function with $z$-dependence. The latter because the expression $\beta = -\tfrac12 ( \widetilde\vartheta + i  \vartheta )$ may be interpreted as a pulse, evaluated at $z=0$, that produces the potential (\ref{pt1}). A first proposal may be to consider the following linear superposition of bright and dark solitons
\be
\Lambda (x;z) = -\tfrac12 \left[ \widetilde\Theta(x;z) +i \Theta(x;z) \right] = \beta(x) e^{i \kappa^2 z}, \quad \Lambda(x; z=0)= \beta(x).
\label{betasol}
\ee
In this case the $\beta$-function works as the stationary form of $\Lambda$, which is factorized by $\beta$ itself and a $z$-dependent phase. Notice that the constraint (\ref{modbeta}) is not affected by the $z$-dependence since $\vert \Lambda(x;z) \vert^2 = \vert \beta(x) \vert^2 =\frac{\kappa^2}{4}$. Then, although the superposition (\ref{betasol}) may describe the $z$-dependence of the $\beta$-function, its intensity does not change shape as the pulse $\Lambda(x;z)$ propagates along the $z$-axis. Nevertheless, a striking expression for $\beta(x)$ is still available. 

\begin{figure}[htb]

\centering
\subfigure[ ]{\includegraphics[width=0.3\textwidth]{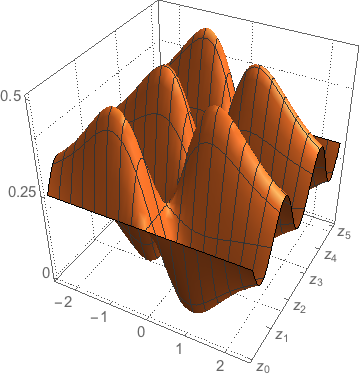}} 
\hspace{1cm}
\subfigure[ ]{\includegraphics[width=0.3\textwidth]{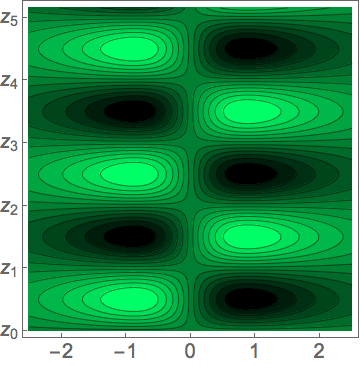}}

\caption{\footnotesize 
Pulse (\ref{betasol3}) generated by the linear superposition (\ref{betasol2}) that includes the bright soliton $\Theta(x;z)$ and the stationary dark soliton $\widetilde\vartheta(x)$. The excitation oscillates as $z$ increases, the constraint $\vert \Gamma (x;z) \vert^2= \sfrac{\kappa^2}{4}$ is satisfied at the points $z_n = (\sfrac{\pi}{\kappa^2}) n$, where the pulse becomes flat.
At $z= \sfrac{z_1}{2}$, the configuration involves a hole (dark soliton) in $x>0$, and a hill (bright soliton) in $x<0$, which acquires a new shape at $z=\sfrac{z_3}{2}$ since it includes a hill in $x>0$, and a hole in $x<0$. 
({\bf a}) The pulse propagates from $z_0$ to $z_5$. ({\bf b}) Distribution of holes and hills along the $z$-axis.
}
\label{Fig1A}
\end{figure}

Let us assume that only the initial value of $\widetilde\Theta(x;z)$ is involved in the definition of the $z$-parameterized version of $\beta$, and that the phase of $\Theta(x;z)$ is permitted at any $z$. In this case we can introduce a different superposition
\be
\Gamma (x;z) = -\tfrac12 \left[ \widetilde\vartheta(x) +i e^{i \kappa^2 z} \vartheta(x) \right] = -\tfrac12 \left[ \widetilde\vartheta(x) - \sin (\kappa^2 z) \vartheta(x) +i \cos (\kappa^2 z) \vartheta(x)
\right],
\label{betasol2} 
\ee
which satisfies $\Gamma(x; z=0) = \beta(x)$ but is not factorized as a function of $x$ times a function of $z$. Remarkably, the intensity
\be
\vert \Gamma (x;z) \vert^2 = \tfrac{\kappa^2}{4} - \tfrac12 \sin (\kappa^2 z) \widetilde\vartheta(x) \vartheta(x) 
\label{betasol3}
\ee
oscillates with period $\frac{ 2\pi}{ \kappa^2}$ as the pulse $\Gamma(x;z)$ propagates along $z$. The constraint (\ref{modbeta}) is satisfied at $\pm z_n = \pm \left( \frac{\pi}{\kappa^2} \right) n$, with $n=0,1,\ldots$ In Fig.~\ref{Fig1A} we can appreciate that the excitation (\ref{betasol3}) is indeed a pair `hole-hill' that borns shyly at $z=0$, maturates up to a robust configuration at $z= \frac{\pi}{2 \kappa^2}
$, and decays slowly up to its annihilation at $z_1=\frac{\pi}{\kappa^2}$. Then the configuration twirls to provide a pair `hill-hole', and the process initiates again to finish at $z_2 = \frac{2\pi}{\kappa^2}$. The entire cycle $z_0 \rightarrow z_2$ is repeated over and over as $z$ grows up. The annihilation positions $z_n$ define a flat configuration of the excitation that serves to construct the potential (\ref{pt1}).

\section{Conclusion}
\label{concluye}

We have shown that the Darboux deformations of the free-particle potential, generated by the soliton-like solutions of the cubic nonlinear Schr\"odinger equation, lead to complex-valued potentials that can be associated with the refractive-index of Kerr media. The real part of these potentials is equal to the intensity of a fundamental bright soliton while the imaginary part is obtained from the product of such a soliton and its dark counterpart. In both cases, the $z$-dependence of the solitons is canceled to produce a $z$-independent potential, with $z$ the propagation direction (which is transversal to the space-coordinate $x$ defining the potential and its solutions). In this form, the real and imaginary parts of the potential refer to the guiding of waves along $z$, and to the balanced gain-and-loss rates of the related intensity, respectively.

The model can be scaled in different directions. For instance, fundamental solitons may be replaced by excited modes in the definition of $\beta$, so it becomes a superposition of excited localized modes of the cubic nonlinear Schr\"odinger equation. Remarkably, the difficulty of using excited physical energies in the Darboux transformation is not present in the construction of complex-valued potentials since the conventional oscillation theorems do not operate in such a case \cite{Jai17} (see also the discussion on the matter in, e.g. \cite{Bla19,Zel01}). Then, it is expected the same situation for the excited soliton modes. Another option trends towards the Gross--Pitaevskii equation where the external potential is not trivially zero. Namely, to satisfy the constraint (\ref{modbeta}) that delimits the class of external potentials $V_0$ that are useful in our model, periodic potentials might be investigated. The same holds for the family of transparent potentials that either vanish or become finite asymptotically. In any case, the complex-valued potential $V = V_0 + 2 \beta_x$ will be expressed as $V=\vartheta^2 +i \vartheta_x$, with $u$ a localized mode of either the Gross--Pitaevskii equation or the cubic nonlinear Schr\"odinger equation.

\appendix
\section{Time-independent Gross--Pitaevskii equation vs Ermakov equation}
\label{ApA}

\renewcommand{\thesection}{A-\arabic{section}}
\setcounter{section}{0}  

\renewcommand{\theequation}{A-\arabic{equation}}
\setcounter{equation}{0}  

In Section~\ref{GPE}, using the relationship between $\beta$ and $\vartheta$ defined in Eq.~(\ref{betas}), from the constraint (\ref{modbeta}) we have derived the nonlinear differential equation (\ref{tetita}), written
\be
\vartheta_x^2 + (4\epsilon - 2 V_0) \vartheta^2 + \vartheta^4 = 0.
\label{A0}
\ee
In similar form we may arrive at the following nonlinear differential equation
\be
\alpha_x^2 + \frac{\lambda^2}{\alpha^2} = \left( \tfrac12 V_0 - \epsilon \right) \alpha^2.
\label{A1}
\ee
As equations (\ref{tetita}) and (\ref{teta4})-(\ref{teta4b}) must be consistent we reduce the power of $\vartheta_x$ in (\ref{A0}) by derivating with respect to $x$, which yields
\be
\vartheta_x [\vartheta_{xx} + 4 \epsilon \vartheta -(V_0 -2 \vartheta^2) \vartheta]= \vartheta (V_0 \vartheta)_x.
\label{A2}
\ee
Considering nontrivial functions $\vartheta$ we conclude that the condition $V_0 \vartheta = \operatorname{const}$ is sufficient to satisfy Eq.~(\ref{teta4}).

In turn, equations (\ref{A1}) and (\ref{Ermakov}) must be also equivalent. In this case the derivative of (\ref{A1}) produces
\be
4\alpha_x \left[\alpha_{xx} - (V_0 -\epsilon) \alpha - \frac{\lambda^2}{\alpha^3} \right] = \alpha [ V_{0x} \alpha - 2 V_0 \alpha_x].
\label{A3}
\ee
As $\alpha$ is positive semidefinite, to satisfy the Ermakov equation (\ref{Ermakov}) we 
make
\be
\frac{d}{dx} \ln \left( \frac{V_0}{\alpha^2} \right) = 0.
\label{A4}
\ee
Thus, $V_0 \alpha^{-2}= \operatorname{const}$. Considering (\ref{sol}) we realize that the latter condition is indeed the same as the one found in the previous case.

The above results suggest a relationship between the time-independent Gross--Pitaevskii equation (\ref{teta4})-(\ref{teta4b}) and the Ermakov one (\ref{Ermakov}). To show that it is the case notice that the successive derivatives of Eq.~(\ref{sol}) yield
\be
\vartheta_x = -2\vartheta \left( \frac{\alpha_x}{\alpha} \right), \quad \vartheta_{xx} = 6\vartheta \left( \frac{\alpha_x}{\alpha} \right)^2 - 2 \vartheta \left( \frac{\alpha_{xx}}{\alpha} \right).
\label{A5}
\ee
Therefore, Eq.~(\ref{teta4}) becomes
\be
-6 \alpha_x^2 + 2 \alpha_{xx} \alpha + V_0 \alpha^2 - 8\frac{\lambda^2}{\alpha^2} = 4 \epsilon \alpha^2.
\label{A6}
\ee
Assuming that the Ermakov equation (\ref{Ermakov}) is fulfilled, the above equation is reduced to (\ref{A1}). That is, the nonlinear equations (\ref{teta4}) and (\ref{Ermakov}) are interrelated through the transformation (\ref{sol}) and the derivative of Eq.~(\ref{A1}) whenever $V_0\alpha^{-2}=\operatorname{const}$.

\section*{Acknowledgments}

We acknowledge the financial support from the Spanish MINECO (Project MTM2014-57129-C2-1-P), Junta de Castilla y Le\'on (VA057U16), Instituto Polit\'ecnico Nacional, Mexico (Project SIP20200818), and Consejo Nacional de Ciencia y Tecnolog\'ia, Mexico (Grant Number A1-S-24569).

\section*{Conflicts of interest}

This work does not have any conflicts of interest.



\begin{thebibliography}{99}

\bibitem{Agr92}
G.P. Agrawal and R.W. Boyd, {\em Contemporary Nonlinear Optics}, Academic Press, New York, 1992.

\bibitem{Tay92}
J.R. Taylor (Ed.), {\em Optical Solitons--Theory and Experiment}, Cambridge University Press, Cambridge 1992.

\bibitem{Kiv03}
Y.S. Kivshar and G. Agrawal, {\em Optical Solitons: From Fibers to Photonic Crystals},
Academic Press, San Diego, California, 2003.

\bibitem{Lam80}
G.L. Lamb, {\em Elements of Soliton Theory}, John Wiley \& Sons, New York, 1980.

\bibitem{Nov84}
S. Novikov, S.V. Manakov, L.P. Pitaevskii and V.E. Zakharov, {\em Theory of Solitons. The Inverse Scattering Method}, Consultants Bureau, New York, 1984.

\bibitem{Kiv98}
Y.S. Kivshar and B. Luther-Davies, Dark optical solitons: physics and applications, {\em Phys. Rep.} {\bf 298} (1998) 81.

\bibitem{Sul99}
C. Sulem and P.L. Sulem, {\em The Nonlinear Schr\"odinger Equation. Self-Focusing and Wave Collapse}, Springer, New York, 1999.

\bibitem{Kom15a}
Y. Kominis, Dynamic power balance for nonlinear waves in unbalanced gain and loss landscapes, {\em Phys. Rev. A} {\bf 92} (2015) 063849.

\bibitem{Kom15b}
Y. Kominis, Soliton dynamics in symmetric and non-symmetric complex potentials, {\em Opt. Commun.} {\bf 334} (2015) 265.

\bibitem{Bis17a}
A. Biswas et al., Resonant optical solitons with quadratic-cubic nonlinearity by semi-inverse variational principle, {\em Optik} {\bf 145} (2017) 18.

\bibitem{Bis17b}
A. Biswas et al., Conservation laws for cubic–quartic optical solitons in Kerr and power law media, {\em Optik} {\bf 145} (2017) 650.

\bibitem{Yan18}
C. Yang et al., Amplification, reshaping, fission and annihilation of optical solitons in dispersion-decreasing fiber, {\em Nonlinear Dyn.} {\bf 92} (2018) 203

\bibitem{Li18}
W. Li et al., Soliton structures in the (1+1)-dimensional Ginzburg--Landau equation with a parity-time-symmetric potential in ultrafast optics, {\em Chin. Phys.} B {\bf 27} (2018) 030504.

\bibitem{Liu18}
M. Liu et al., Ultrashort pulse generation in mode-locked erbium-doped fiber lasers with tungsten disulfide saturable absorber, {\em Optics Comm.} {\bf 406} (2018) 72.

\bibitem{Yu18}
W. Yu et al., Periodic oscillations of dark solitons in nonlinear optics, {\em Optik} {\bf 165} (2018) 341.

\bibitem{Guo18}
H. Guo et al., Analytic study on interactions of some types of solitary waves, {\em Optik} {\bf 164} (2018) 132.

\bibitem{Zha18}
Y. Zhang et al., Some types of dark soliton interactions in inhomogeneous optical fibers, {\em Opt. Quant. Electron.} {\bf 50} (2018) 295.

\bibitem{Rut10}
C.E. R\"uter et al., Observation of parity-time symmetry in optics, {\em Nat. Phys.} {\bf 6} (2010) 192.

\bibitem{Ben05}
C.M. Bender, Introduction to PT-symmetric quantum theory, {\em Contemp. Phys.} {\bf 46} (2005) 277.

\bibitem{Ros15}
O. Rosas-Ortiz, O Casta\~nos, and D. Schuch, New supersymmetry-generated complex potential with real spectra, {\em J. Phys. A: Math. Theor.} {\bf 48} (2015) 445302.

\bibitem{Zel16}
K.D. Zelaya and O. Rosas-Ortiz, Optimized Binomial Quantum States of Complex Oscillators with Real Spectrum, {\em J. Phys.: Conf. Ser.} {\bf 698} (2016) 012026.

\bibitem{Jai17}
A. Jaimes-Najera and O. Rosas-Ortiz, Interlace properties for the real and imaginary parts of the wave functions of complex-valued potentials with real spectrum, {\em Ann. Phys.} {\bf 376} (2017) 126.

\bibitem{Ros18}
O. Rosas-Ortiz and K. Zelaya, Bi-Orthogonal Approach to Non-Hermitian Hamiltonians with the Oscillator Spectrum: Generalized Coherent States for Nonlinear Algebras, {\em Ann. Phys.} {\bf 388} (2018) 26.

\bibitem{Bla19}
Z. Blanco-Garcia, O. Rosas-Ortiz and K. Zelaya, Interplay between Riccati, Ermakov and Schr\"odinger equations to produce complex-valued potentials with real energy spectrum, {\em Math. Meth. Appl. Sci.} {\bf 42} (2019) 4925.

\bibitem{Zel01}
K. Zelaya, S. Cruz y Cruz and O. Rosas-Ortiz, On the construction of non-Hermitian Hamiltonians with all-real spectra through supersymmetric algorithms, arXiv:2001.02794

\bibitem{Hil97}
E. Hille, {\em Ordinary Differential Equations in the Complex Domain}, Dover, New York, 1997.

\bibitem{Sch18}
D. Schuch, {\em Quantum Theory from a Nonlinear Perspective, Riccati Equations in Fundamental Physics}, Springer, Berlin, 2018.

\bibitem{Erm80}
V. Ermakov, Second order differential equations. Conditions of complete integrability, {\em Kiev University Izvestia}, Series III 9 (1880) 1 (in Russian). English translation by A.O. Harin, {\em Appl. Anal. Discrete Math.}{\bf 2} (2008) 123. 


\bibitem{Dar82}
G. Darboux, Sur une proposition relative aux \'equations lin\'eares, {\em C.R. Acad. Sci. Paris} {\bf 94} (1882) 1456.

\bibitem{Mie04}
B. Mielnik and O. Rosas-Ortiz, Factorization: Little or great algorithm?, {\em J. Phys. A: Math. Gen.} {\bf 37} (2004) 10007.

\bibitem{Rog02}
C. Rogers and W.K. Schief, {\em B\"acklund and Darboux Transformations. Geometry and Modern Applications in Soliton Theory}, Cambridge University Press, United Kingdom, 2012.

\bibitem{Coo01}
F. Cooper, A. Khare and U. Sukhatme, {\em Supersymmetry in Quantum Mechanics},  World Scientific, Singapore, 2001.

\bibitem{Gro61}
E.P. Gross, Structure of a quantized vortex in boson systems, {\em  Il Nuovo Cimento} {\bf 20} (1961) 454.

\bibitem{Pit61}
L.P. Pitaevskii, Vortex lines in an imperfect Bose gas, {\em  Sov. Phys. JETP} {\bf 13} (1961) 451.

\bibitem{Ele17}
H. Eleuch and I. Rotter, Gain and loss in open quantum systems, {\em Phys. Rev. E} {\bf 95} (2017) 062109.

\bibitem{Rog13}
J. Rogel-Salazar, The Gross-Pitaevskii equation and Bose-Einstein condensates, {\em Eur. J. Phys.} {\bf 34} (2013) 247.

\bibitem{Kev08}
P.G. Kevrekidis, D.J. Frantzeskakis and R. Carretero-Gonz\'alez, {\em Emergent Nonlinear Phenomena in Bose-Einstein Condensates. Theory and Experiment}, Springer, Berlin, 2008.

\bibitem{Koh59}
W. Kohn, Analytic Properties of Bloch Waves and Wannier Functions, {\em Phys. Rev.} {\bf 115} (1959) 809.

\bibitem{Tro01}
A. Trombettoni and A. Smerzi, Discrete Solitons and Breathers with Dilute Bose-Einstein Condensates, {\em Phys. Rev. Lett.} {\bf 86} (2001) 2353.

\bibitem{Zak71}
V.E. Zakharov and A.B. Shabat, Exact theory of two-dimensional self-focusing and one-dimensional self-modulation of waves in nonlinear media, {\em Zh. Eksp. Teor. Fiz.} {\bf 61} (1971) 118; {\em Sov. Phys. JETP} {\bf 34} (1972).

\bibitem{Dia99}
J. I. D\'{\i}az, J. Negro, L. M. Nieto and O. Rosas-Ortiz, The supersymmetric modified P\"oschl-Teller and delta-well potentials, {\em J. Phys. A: Math. Gen.} {\bf 32} (1999) 8447.

\bibitem{Mie00}
B. Mielnik, L.M. Nieto and O. Rosas-Ortiz, The finite difference algorithm for higher order supersymmetry, {\em Phys. Lett.} {\bf A 269} (2000) 70.

\bibitem{Lev00}
G. Levai and M. Znojil, Systematic search for $PT$-symmetric potentials with real energy spectra, {\em J. Phys. A: Math. Gen.} {\bf 33} (2000) 7165.

\bibitem{Mus08}
Z. H. Musslimani, K. G. Makris, R. El-Ganainy, and D. N. Christodoulides, Optical Solitons in ${\cal PT}$ Periodic Potentials, {\em Phys. Rev. Lett.} {\bf 100} (2008) 030402.

\bibitem{Mid14}
B. Midya and R. Roychoudhury, Nonlinear localized modes in ${\cal PT}$-symmetric optical media with competing gain and loss, {\em Ann. Phys.} {\bf 341} (2014) 12.

\bibitem{Tso14}
E.N. Tsoy, I.M. Allayarov and F. Kh. Abdullaev, Stable localized modes in asymmetric waveguides with gain and loss, {\em Opt. Lett.} {\bf 39} (2014) 4215

\bibitem{Che14}
H. Chen, D. Hu and L. Qi, The optical solitons in the Scarff parity-time symmetric potentials, {\em Opt. Comm.} {\bf 331} (2014) 139.

\bibitem{Ahm01}
Z. Ahmed, Real and complex discrete eigenvalues in an exactly solvable one-dimensional complex ${\cal PT}$-invariant potential, {\em Phys. Lett.} {\bf A 282} (2001) 343.


\end{thebibliography}
\end{document}